\documentclass[11pt]{article}

\usepackage[utf8]{inputenc}
\usepackage[T1]{fontenc}
\usepackage{amsmath,amssymb,amsthm}
\usepackage{graphicx}
\usepackage{geometry}
\usepackage{hyperref}
\hypersetup{
    colorlinks=true,
    linkcolor=blue,
    citecolor=blue,
    urlcolor=blue
}
\usepackage[numbers,sort&compress]{natbib}   
\usepackage{booktabs}      
\usepackage{caption}
\usepackage{subcaption}
\usepackage{xcolor}
\usepackage{lipsum} 
\usepackage[version=4]{mhchem}
\usepackage{siunitx}
\usepackage{graphicx}%
\usepackage{multirow}%
\usepackage{amsmath,amsfonts}%
\usepackage{amsthm}%
\usepackage{mathrsfs}%
\usepackage[title]{appendix}%
\usepackage{xcolor}%
\usepackage{textcomp}%
\usepackage{manyfoot}%
\usepackage{booktabs}%
\usepackage{algorithm}%
\usepackage{algorithmicx}%
\usepackage{algpseudocode}%
\usepackage{listings}%
\usepackage{authblk}

\theoremstyle{definition}

\theoremstyle{remark}

\title{State-dependent recruitment of adhesion molecules enables perfect stabilization in cell-adhesion models}
\author[1,2,3]{Anton F. Burnet}
\author[2]{Julia Müllner}
\author[1,3,*]{Benedikt Sabass}
\affil[1]{Department of Physics, Technische Universit\"at Dortmund, Dortmund, 44227, Germany}
\affil[2]{Faculty of Physics and Center for NanoScience, Ludwig-Maximilians-Universit\"at M\"unchen, M\"unchen, 80752, Germany}
\affil[3]{Department of Veterinary Sciences, Ludwig-Maximilians-Universit\"at M\"unchen, M\"unchen, 80752, Germany}
\affil[*]{benedikt.sabass@tu-dortmund.de}

\date{\today}

\begin{document}

\maketitle

\begin{abstract}
Cells adapt their adhesion to mechanical load, but the physical conditions under which this response prevents rupture remain unclear. Inspired by focal adhesions, we study models in which force-sensitive conformational states within an adhesion cluster are coupled to the recruitment of additional molecules. We show that sufficiently strong coupling creates a distinct regime under load in which clusters grow in proportion to the applied force while the average load per bond remains below the level that destabilizes the cluster. The clusters can therefore withstand arbitrarily large stationary forces in principle. We term this behavior ``perfect stabilization''. At still stronger coupling, the same feedback causes unbounded growth already at equilibrium. For the minimal model, we derive state diagrams that characterize adhesion stability under stationary and dynamic loading. More broadly, using generic molecular-state networks, we derive conditions under which perfect stabilization and related growth instabilities arise in nonequilibrium adhesion systems with state-dependent recruitment.
\end{abstract}

\section{Introduction}
\begin{figure*}[hbt!]
    \centering
\includegraphics[width=.85\linewidth]{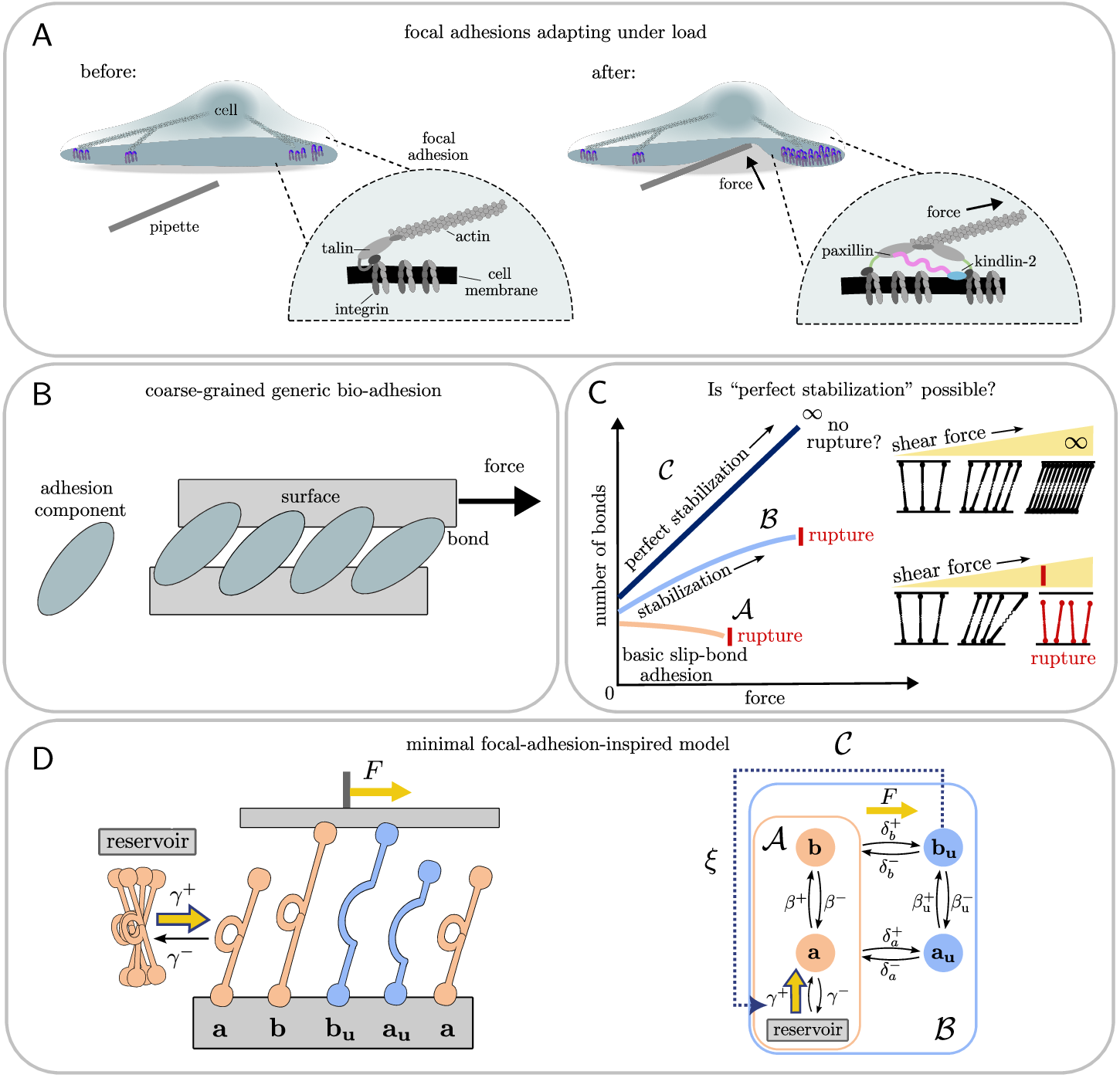} 
    \caption{Focal adhesion growth under shear stress inspires a minimal biomolecular adhesion model for robust mechanosensitive adaptation. (\textsf{A})~Schematic of shear force applied via a pipette to a focal adhesion site of an adherent cell, based on an experiment performed in Ref.~\citep{riveline2001focal}. After application, the focal adhesion site grows proportionally to the load. Enlarged structures show a simplified illustration of talin connecting integrins to actin. As proposed in Ref.~\citep{lu2022mechanism}, activated talin can dimerize and interact through a paxillin–kindlin-2 complex to promote integrin clustering. (\textsf{B})~Schematic of a generic bio-adhesion structure under force, coarse-grained into components (e.g., proteins or filaments), which form bonds between surfaces. (\textsf{C})~A basic slip-bond adhesion system typically destabilizes under load, resulting in a decrease in the number of bonds (orange curve, subgraph $\mathcal{A}$ in D). Self-stabilizing adhesion clusters exhibit an increase in bond numbers with applied force up to a critical point where bond-rupturing events dominate~\citep{braeutigam2022generic} (light blue curve, subgraph $\mathcal{B}$ in D). Perfect stabilization entails a sustained increase in bond number with force, preventing rupture by keeping the average molecular stretch low (dark blue curve, $\mathcal{C}$ in D). (\textsf{D})~A minimal system for modeling basic slip bonds (subgraph $\mathcal{A}$), as well as self-stabilization through 
   force-sensitive unfolding of molecules (subgraph $\mathcal{B}$). We furthermore introduce a coupling $\xi$ ($\mathcal{C}$, dotted arrow) between force-sensitive molecule unfolding and molecule recruitment.}
    \label{fig:main}
\end{figure*}

Cell adhesions are dynamic molecular assemblies that transmit forces between the cytoskeleton and the extracellular environment. Focal adhesions (FAs) are prominent cell–matrix adhesion structures that play important roles in mechanosensing~\citep{geiger2009environmental,schoen2013yin}, cell migration~\citep{huttenlocher2011integrins}, development~\citep{engler2006matrix,heisenberg2013forces}, and disease~\citep{butcher2009tense, schwager2019cell, xin2023biophysics}. Unlike conventional adhesives, which typically weaken under sustained load, focal adhesions can increase in size in response to externally applied shear force~\citep{riveline2001focal} (see Fig. \ref{fig:main}\textsf{A}). This response involves the recruitment of additional adhesion molecules and growth of the adhesive contact. Which biophysical mechanisms enable force-induced adhesion growth, and what are the force and loading-rate limits of the resulting stabilization?

The architecture of FAs comprises transmembrane integrins that bind the extracellular matrix (ECM) and connect intracellularly to adaptor proteins that dynamically associate and dissociate with force-generating actomyosin filaments of the cytoskeleton. This architecture has led to the characterization of FAs as molecular clutches, where the integrin-adaptor-protein complex constitutes the clutch unit~\citep{mitchison1988cytoskeletal, gardel2008traction, sabass2010modeling, li2010model, moore2010stretchy, sens2013rigidity,sun2016integrin, chan2008traction,liu2015talin, elosegui2018control, ron2020one, driscoll2020actin, liu2025elastic, sens2020stick, heyn2025cell, novikova2021evolving}. Among the adaptor proteins~\citep{zaidel2003early}, talin plays a principal role in mechanotransduction~\citep{lee2007force, yan2015talin,kumar2016talin,klapholz2017talin,goult2018talin}. Central to talin’s mechanosensitivity is its ability to unfold under strain~\citep{del2009stretching}. From a biological perspective, however, talin's function extends beyond simply unfolding. Force-induced talin unfolding exposes cryptic vinculin-binding sites, thereby promoting additional molecular interactions within the adhesion~\citep{del2009stretching,yao2014mechanical,yao2016mechanical}. Talin-dependent interactions have also been implicated in integrin clustering and further recruitment of talin~\citep{cluzel2005mechanisms, roca2009clustering, litschel2024membrane}. Related force-dependent reinforcement mechanisms have been invoked to explain focal-adhesion adaptation to extracellular-matrix stiffness~\citep{elosegui2016mechanical} and dynamic loading~\citep{andreu2021force}. The molecular signaling pathways connecting talin-based mechanotransduction to integrin clustering remain incompletely understood. For example, it was recently shown that talin molecules dimerize through interactions of their C-terminal rod domains and bind via paxillin to integrin-bound kindlin-2, forming integrin-activation complexes that further enhance integrin clustering~\citep{lu2022mechanism} (see Fig. \ref{fig:main}\textsf{A}). Overall, however, experimental data support the hypothesis that the conformational or binding states of molecules already present in an adhesion can regulate the recruitment of further components. 

To navigate the biological complexity of FAs, theoretical work has sought to distill core physical principles behind adhesion growth~\citep{ladoux2012physically,schwarz2013physics}. Prominent approaches include thermodynamic descriptions~\citep{nicolas2004cell,shemesh2005focal,besser2006force,nicolas2008dynamics,smith2008force,olberding2010non} and modeling of the stochastic bond dynamics with so-called ``clutch models''~\citep{kong2010stabilizing,gao2011probing,de2018general,braeutigam2022generic,braeutigam2024clutch} (see Fig.~\ref{fig:main}\textsf{B}). In a basic slip-bond cluster, bond lifetimes decrease with force, causing the adhesion to shrink and eventually rupture under increasing load~\citep{merkel1999energy,jiang2003two}, see subgraph~$\mathcal{A}$ in Fig.~\ref{fig:main}(\textsf{C}, \textsf{D}). Catch-bond kinetics~\citep{huang2017vinculin,chen2017force,Owen} can increase individual bond lifetimes over particular force ranges, but their role in recruitment-driven growth of the entire adhesion cluster remains unclear. Inspired by studies of talin-based mechanotransduction~\citep{elosegui2016mechanical}, modeling work revealed that slip-bond adhesion clusters can grow under increasing load if the molecules can unfold and are exchanged with a reservoir~\citep{braeutigam2022generic,braeutigam2024clutch}, see subgraph~$\mathcal{B}$ in Fig.~\ref{fig:main}(\textsf{C}, \textsf{D}). This mechanism, termed ``self-stabilization'', occurs up to a limiting force before total rupture. It remains unclear whether additional physical mechanisms could eliminate finite rupture thresholds altogether.

In this work, we study a class of biomolecular adhesion models in which the recruitment of load-bearing molecules depends on the states of molecules already present in the cluster. The internal molecular transitions obey local detailed-balance constraints at equilibrium. We begin with an established adhesion model that includes force-sensitive molecular unfolding and exchange with a reservoir (subgraph $\mathcal{B}$, Fig.~\ref{fig:main}\textsf{D}), and introduce a coupling between the unfolded, load-bearing state and the recruitment of additional molecules (subgraph $\mathcal{C}$, Fig.~\ref{fig:main}\textsf{D}). This coupling produces a qualitatively distinct regime in which adhesion growth prevents rupture at any finite stationary force within the model. We term this regime ``perfect stabilization'' because, within the model, no finite stationary force causes rupture. The availability of an unlimited adhesion-molecule reservoir is not sufficient for perfect stabilization. Both the basic slip-bond model and the previously described self-stabilizing model can comprise a molecule reservoir but nevertheless exhibit finite rupture thresholds. Perfect stabilization specifically requires state-dependent recruitment to generate a growth instability at a mechanically accessible, finite load per bond while the zero-load state remains stable. For a minimal model, we characterize the transitions between limited self-stabilization, perfect stabilization, and unbounded growth under stationary and dynamic loading. We then derive general conditions for perfect stabilization in molecular-state networks with load-dependent recruitment.

\section{Results}
\subsection{Model}
An adhesion site is conceptualized as two parallel surfaces that are held together by a cluster of adhesion molecules, see Fig.~\ref{fig:main}\textsf{D}. A shear force $F$ acts on the upper plane and induces a displacement $s$ and an average velocity $v = \langle\dot{s}\rangle$. For simplicity, all adhesion molecules are modeled as Hookean springs with spring constant $\kappa$ and stretch $h$. Furthermore, it is assumed that viscous forces are negligible so that mechanical relaxation occurs instantaneously.
\\
We first focus on a basic slip-bond adhesion site consisting only of the states $a$ and $b$, highlighted as case $\mathcal{A}$ in Fig.~\ref{fig:main}\textsf{D}. State $a$ represents molecules that are only connected to the lower surface. Molecules coming from the surrounding bulk, termed reservoir, assemble at the adhesion site with rate $\gamma^+$. Dissociation of molecules from the cluster is governed by a rate coefficient $\gamma^-$. In state $a$, the molecule extension fluctuates freely with variance $\sigma^2 = k_\mathrm{B} T / \kappa$, where $k_\mathrm{B} T$ is the thermal energy. These molecules stochastically form a bond with the upper surface, denoted by state $b$. The binding transition $a \rightarrow b$ is governed by the rate coefficient $\beta^+(h) = k_{\beta} \exp\left(-\frac{\left(|h| - \ell_\mathrm{b} \right)^2}{2 \sigma^2} +\frac{h^2}{2\sigma^2}  + \frac{\epsilon_\mathrm{b}}{k_\mathrm{B} T} \right)$, where $k_{\beta}$ is the intrinsic binding rate, $\epsilon_\mathrm{b}$ an effective binding affinity, and $\ell_\mathrm{b}$ an optimal binding distance. The dissociation of a bound molecule $b \rightarrow a$ is governed by the rate coefficient $\beta^-(h) = k_{\beta} \exp\left(\frac{2|h|\ell_\mathrm{b} - \ell_\mathrm{b}^2}{2 \sigma^2} \right)$. The frequency of unbinding events increases with load, causing a monotonic reduction in bonds with shear force $F$ in this basic model, until the cluster abruptly undergoes total rupture, see case $\mathcal{A}$ in Fig.~\ref{fig:main}\textsf{C}.
\\
The basic slip-bond model can be extended to include an unfolded molecule conformation, see case $\mathcal{B}$ in Fig.~\ref{fig:main}\textsf{D}. We designate states and transition rates pertaining to unfolded molecules with a subscript $\mathrm{u}$. This simple model extension can produce adhesion clusters that, as a whole, display a fundamentally different force response than individual bonds. Here, stretch shifts the state occupations to drive growth of the adhesion cluster~\citep{braeutigam2022generic}. Growth stabilizes adhesion clusters reversibly under force, in spite of the bond properties being the same as in the basic slip-bond model.
This self-stabilization occurs up to a point of total rupture, see Fig.~\ref{fig:main}\textsf{C}. Molecule unfolding is modeled as a thermally assisted jump over a single energy barrier, yielding a rate coefficient of $\delta_{a,b}^+(h)= k_\delta \exp\left( \frac{2 \Delta_{1} h - \Delta_{1}^2}{2\sigma^2}  -\frac{\epsilon_\mathrm{f}}{k_\mathrm{B}T} \right)$ while refolding follows $\delta_{a,b}^-(h)= k_\delta \exp\left( \frac{-2 \Delta_2 h -  \Delta_2^2}{2\sigma^2}  \right)$, where $\Delta_{1,2}$ are the distances between the metastable energy minima of the folded and unfolded states and the barrier maximum, $\epsilon_\mathrm{f}$ is a constant energy required for the conformation change, and $k_{\delta}$ is an intrinsic rate, where we take $k_{\delta}=k_{\beta}$. The unfolding length is given by $\Delta=\Delta_1+\Delta_2$ and we assume $\Delta_1=\Delta_2=\Delta/2$. Hence, upon unfolding the stretch $h$ relaxes to $h-\Delta$. All rates are defined such that the system is thermodynamically consistent at equilibrium (Methods and Supplementary Information). 
\\
Motivated by evidence that activated or force-sensitive states of resident focal-adhesion proteins can promote further molecular recruitment, we introduce state-dependent recruitment into the model. We model this type of recruitment by making the rate $\gamma^+$ depend on the number of bound and unfolded molecules that are already in the cluster $N_{b_\mathrm{u}}$ as $\gamma^+=\gamma_0^+ + \xi N_{b_\mathrm{u}}$, where $\gamma^+_0, \,\xi\in\mathbb{R}^+$. See case $\mathcal{C}$ in Fig.~\ref{fig:main}\textsf{D}. This seemingly simple modification has a profound impact on the adhesion stability, as elaborated in the following.

\subsection{Emergence of perfect stabilization}

\begin{figure}[hbt!]
    \centering
  \includegraphics[width=0.6\linewidth]{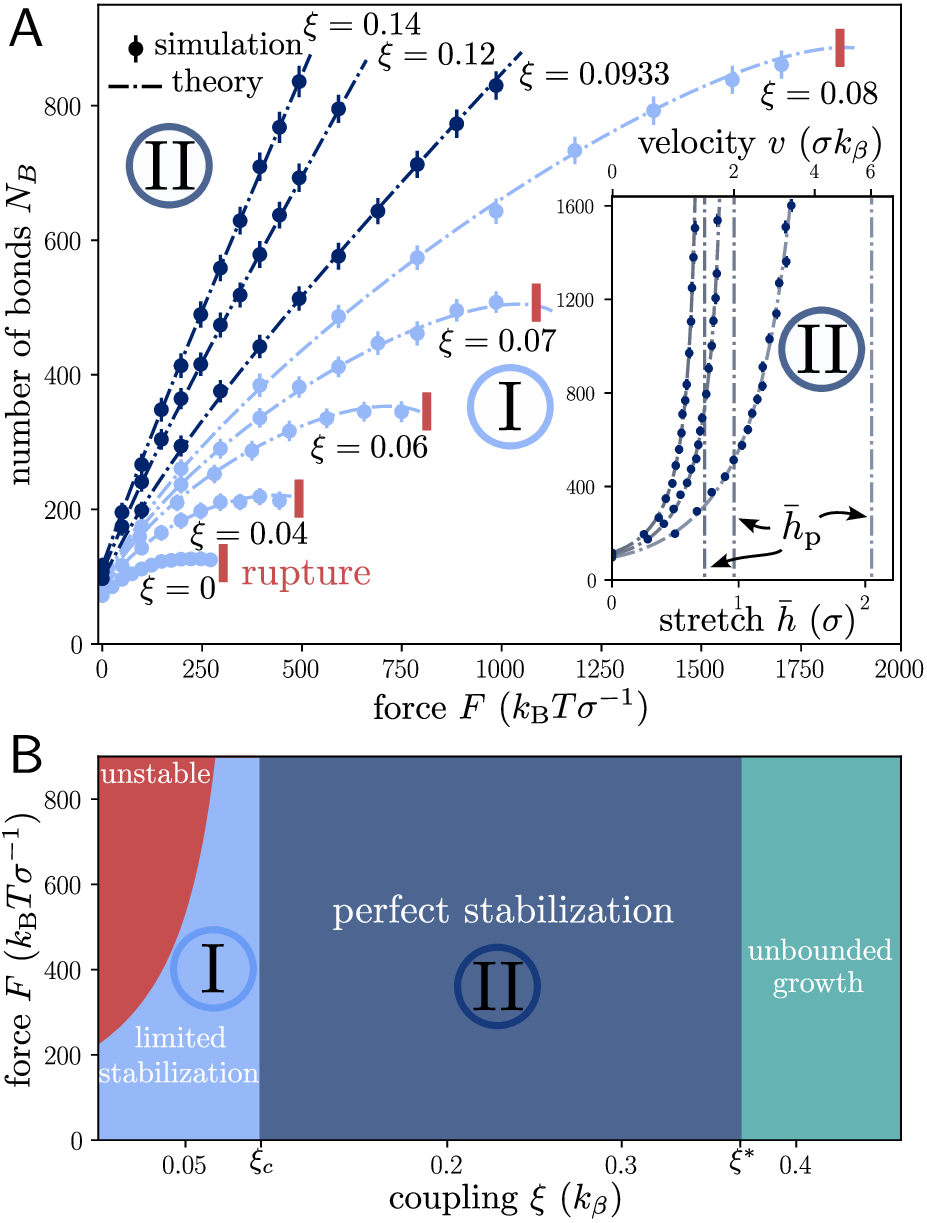} 
    \caption{Simulation results for the model shown in Fig.~\ref{fig:main}\textsf{D}. Dashed lines show mean-field solutions. (\textsf{A})~For small coupling $\xi$, the mean number of bound molecules in steady state $N_B$ initially grows with force up to a point of saturation (light blue). Red bars indicate a complete rupture of adhesion clusters in simulations.
    With  $\xi$ increasing, the rupture forces increase. For $\xi\geq\xi_c\approx0.0933k_{\beta}$, adhesion sites no longer rupture in the simulations (dark blue). Inset of (\textsf{A}): $N_B$ plotted against $\Bar{h}$ and $v$ for $\xi\in\{0.0933,0.12,0.14\}k_{\beta}$. $N_B$ diverges for large loads at finite average stretches $\Bar{h}=\Bar{h}_\mathrm{p}$. The error bars represent the SD.
    (\textsf{B})~State diagram for adhesion stability. For $0\leq \xi < \xi_c$, the adhesion clusters self-stabilize in a finite range of forces. For $\xi_c\leq \xi < \xi^*$, the model stabilizes perfectly. For $\xi\geq \xi^*$, the system grows indefinitely.
    }
    \label{fig:sim-full}
\end{figure}
We first perform stochastic simulations of the model (Methods and Supplementary Information). Hereafter, ${N}_{q}$ denotes the steady-state mean number of molecules in state $q$. Figure~\ref{fig:sim-full}\textsf{A} shows the number of all bound molecules ${N}_B={N}_b+{N}_{b_\mathrm{u}}$ as a function of force $F$ for varying coupling strength $\xi$. For small coupling, we observe limited self-stabilization as previously described~\citep{braeutigam2022generic}, where ${N}_B$ initially grows with $F$, followed by a point of saturation and, ultimately, complete rupture. This regime of limited self-stabilization is labeled (\MakeUppercase{\romannumeral 1}) in Fig.~\ref{fig:sim-full}. Remarkably, from a critical value $\xi\geq \xi_c$ on, the adhesion cluster size no longer saturates with increasing load and rupture is not observed in simulations. The inset in Fig.~\ref{fig:sim-full}\textsf{A} illustrates that for $\xi\geq \xi_c$, ${N}_B$ versus the average bond stretch $\Bar{h}$ approaches an asymptote with increasing force, denoted $\Bar{h}_\mathrm{p}$. This asymptotic behavior implies, together with force balance $F=\kappa\Bar{h} N_B$, that ${N}_B$ will eventually grow proportionally to $F$ while the average stretch attains a finite constant. We refer to this regime as \textit{perfect} stabilization, labeled (\MakeUppercase{\romannumeral 2}) in Fig.~\ref{fig:sim-full}. The critical value $\xi_c$ corresponds to the coupling at which the rupture force in regime (\MakeUppercase{\romannumeral 1}) reaches infinity. An increase of $\xi$ beyond $\xi_c$ results in stronger cluster growth with force, maintaining smaller average stretches, until the asymptote $\Bar{h}_\mathrm{p}$ reaches zero for coupling $\xi^*$, marking the onset of an unbounded growth regime, where the system grows indefinitely. The regime of unbounded growth occurs for any force $F \geq 0$. Thus, the perfect stabilization emerges in the bounded interval $\xi_c\leq\xi<\xi^*$. The different regimes are summarized in Fig.~\ref{fig:sim-full}\textsf{B}.

\subsection{Mean-field approximation}
To understand the critical behavior of the adhesion system illustrated in Fig.~\ref{fig:sim-full}, it is sufficient to consider an approximate model in which we decouple the stretch variables of individual bonds from each other by assuming, instead of a constant force $F$, a constant velocity $v$. We describe an ensemble of molecules by functions $n_q(h)$, representing the number of molecules in a state $q$ with stretch $h$. The governing equations for $q,p \in\{a,a_\mathrm{u},b,b_\mathrm{u}\}$ at steady state read
\begin{equation}\label{mean-field-equs}
\begin{split}
    &\partial_t n_q(h)= 0 = \sum_p \left[k^{p\to q}(h)n_p(h) - k^{q\to p}(h)n_q(h)\right]\\ &- (\delta_{q, b }+\delta_{q, b_{\mathrm{u}}}) v \, \partial_h n_q(h) +\delta_{q,a}\,\eta(h)(\gamma_0^+ +\xi {N}_{b_\mathrm{u}})-\delta_{q,a}\gamma^- n_q(h),
\end{split}
\end{equation}
where $k^{p\to q}$ denotes the transition rates from state $p$ to state $q$ in the scheme shown in Fig.~\ref{fig:main}\textsf{D} and $\eta(h)$ denotes the normalized Gaussian distribution $\eta(h)\sim\mathcal{N}(0,\sigma^2)$, where for $q\in\{a, a_\mathrm{u}\}$, $n_q(h)={N}_q\,\eta(h)$. Note that the unfolded molecule states have a shifted stretch coordinate $h-\Delta$. For the bound molecules $q\in\{b, b_\mathrm{u}\}$, the right-hand side of Eq.~\ref{mean-field-equs} has the expression $-v\,\partial_h n_q(h)$ to account for bond stretch due to the relative motion of the planes. Accordingly, the average stretch is a function of velocity $\Bar{h}=\Bar{h}(v)$ and, due to linearity, is independent of $\xi$ (Supplementary Information and Fig.~S1). For the following, we introduce convenient stretch-averaged effective rates as 
\begin{equation}\label{effective_rates}
\bar{k}^{p \to q}=\int_{\mathbb{R}}\mathrm{d}h \, \frac{n_{p}(h)}{{N}_{p}} k^{p\to q}(h).
\end{equation}

\subsection{Instability of equilibrium solution above $\xi^{*}$}
The approximate model becomes exact at equilibrium, where $F=0$, $v=0$, and $\Bar{h}=0$. Since molecule extensions here obey Gaussian distributions, the effective rates are easily calculated. The equilibrium number of molecules in state $q$ is ${N}^{*}_q(\xi) \propto 1/(\gamma^- k^{\to a} - \xi k^{\to b_\mathrm{u}})$, where $k^{\to q}\equiv \sum_\mu \prod_{(p',q')\in \mathcal{T}_\mu^q}\Bar{k}^{p'\to q'}$, which denotes the sum of products of rates $\Bar{k}^{p'\to q'}$ associated with the edge $(p',q')$ that belongs to the $\mu$-th spanning tree of the network graph, shown in Fig.~\ref{fig:main}\textsf{D}, with edges directed towards $q$, $\mathcal{T}_\mu^q$. At $\xi^*\equiv\gamma^- \frac{k^{\to a}}{k^{\to b_\mathrm{u}}}=\gamma^- e^{-(\epsilon_\mathrm{b}-\epsilon_\mathrm{f})/k_\mathrm{B}T}$, the system size diverges as a result of a flux imbalance at the reservoir. Note that the steady-state probability $p_q\propto k^{\to q}$ (Supplementary Information). Therefore, the term $\gamma^-\frac{k^{\to a}}{k^{\to b_\mathrm{u}}}=\gamma^-\frac{p_a}{p_{b_\mathrm{u}}}$. Indeed, by local detailed balance, at equilibrium the steady-state probability ratio follows a Boltzmann distribution $\frac{p_a}{p_{b_\mathrm{u}}}=e^{-(\epsilon_\mathrm{b}-\epsilon_\mathrm{f})/k_\mathrm{B}T}$ where $\epsilon_\mathrm{b}-\epsilon_\mathrm{f}$ corresponds to the free energy difference between the $a$ and $b_\mathrm{u}$ molecule states.

\subsection{Perfect stabilization under load above $\xi_c$}
\begin{figure}[t!]
    \centering
   \includegraphics[width=.6\linewidth]{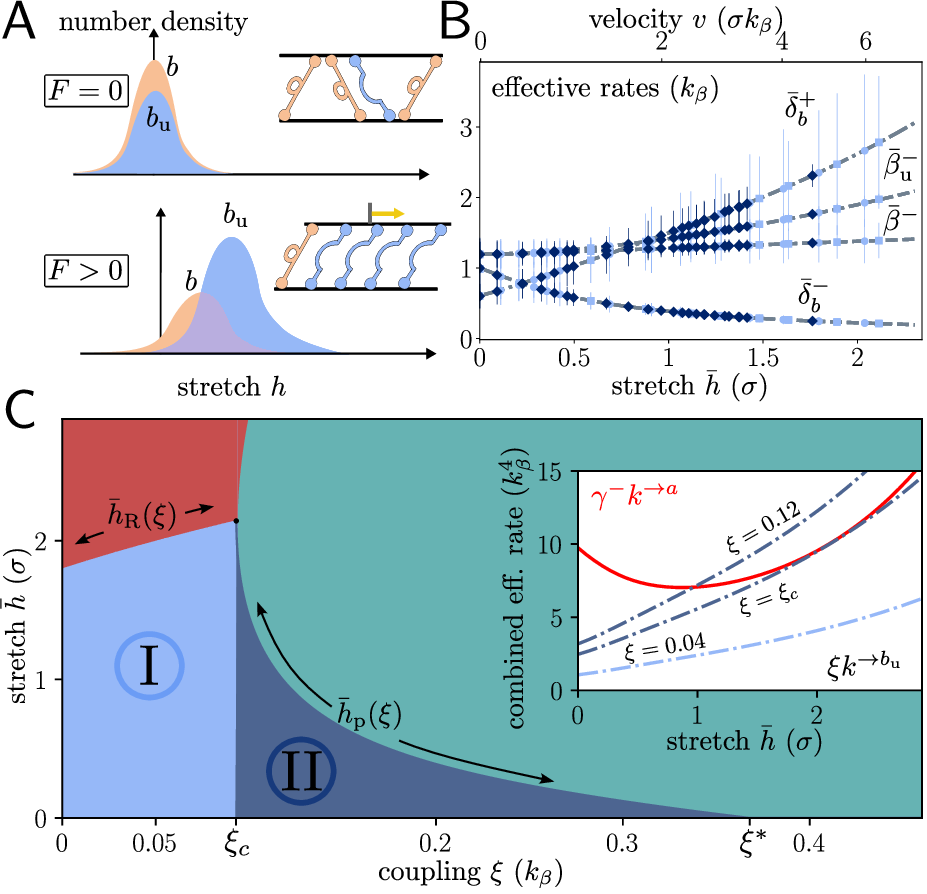}
   \caption{ Model behavior under load, characterized by the average bond stretch. (\textsf{A})~Illustration of the molecule stretch distributions in and out of equilibrium. Forces on the adhesion clusters stretch the bonds, which in turn promotes unfolding of the molecules. (\textsf{B})~Effective transition rates as a function of average stretch $\bar{h}$. Dashed lines are numerical results from the mean-field equations, symbols are results from stochastic simulations (for various $\xi$, both non-critical and critical). The error bars represent the SD. 
   (\textsf{C})~State diagram for adhesion stability as a function of average stretch $\Bar{h}$, corresponding to Fig.~\ref{fig:sim-full}\textsf{B}. Perfect stabilization occurs in region (\MakeUppercase{\romannumeral 2}), where $\xi_c\leq \xi < \xi^*$ and $\Bar{h}<\Bar{h}_\mathrm{p}(\xi)$. The cluster grows without bound on the line $\Bar{h} = \Bar{h}_\mathrm{p}(\xi)$. The inset illustrates the competition between $\gamma^- k^{\to a}$ and $\xi k^{\to b_\mathrm{u}}$ (defined in the main text) with increasing $\Bar{h}$ for varying $\xi$. The points at which the curves cross 
   determine average stretch values $\Bar{h}_\mathrm{p}$ for onset of divergent growth as $\gamma^- k^{\to a}=\xi k^{\to b_\mathrm{u}}$.}
    \label{fig:phase-plot}
\end{figure}

Under load, the bond stretch distributions deviate from a Gaussian and shift in the direction of the applied force, resulting in a non-zero average stretch $\bar{h}$, as depicted in Fig.~\ref{fig:phase-plot}\textsf{A}. 
Due to linearity, the normalized stretch distributions $n_{b_{(\mathrm{u})}}/{N}_{b_{(\mathrm{u})}}$, which can be solved numerically from the mean-field equations (Eq.~\ref{mean-field-equs}) for a given $\Bar{h}$, are independent of $\xi$ (Supplementary Information and Fig.~S1). Consequently, the four $\Bar{h}$-dependent effective rates $\bar{\beta}^-(\Bar{h}),\bar{\beta}_\mathrm{u}^-(\Bar{h}),\bar{\delta}_b^+(\Bar{h})$ and $\bar{\delta}_b^-(\Bar{h})$ are also independent of $\xi$. These effective rates encapsulate the information of the shifting distributions governed by Eq.~\ref{mean-field-equs}, see Fig.~\ref{fig:phase-plot}\textsf{B}. The mean number of bound molecules is therewith given by
\begin{equation}\label{NB-non-equ}
     {N}_B(\Bar{h},\xi) = \frac{\gamma_0^+[k^{\to b}(\Bar{h}) + k^{\to b_\mathrm{u}}(\Bar{h})]}{\gamma^- k^{\to a}(\Bar{h}) - \xi k^{\to b_\mathrm{u}}(\Bar{h})}.
\end{equation}
Solutions of Eq.~\ref{NB-non-equ} are in excellent agreement with simulation results, as shown in Fig.~\ref{fig:sim-full}\textsf{A}. 

To see how Eq.~\ref{NB-non-equ} explains perfect stabilization in the range $\xi_c\leq\xi< \xi^*$, consider regime (\MakeUppercase{\romannumeral 2}) in the state diagram shown in Fig.~\ref{fig:phase-plot}\textsf{C}, corresponding to (\MakeUppercase{\romannumeral 2}) in Fig.~\ref{fig:sim-full}\textsf{B}. The line denoted by $\Bar{h}_\mathrm{p}(\xi)$, where $\xi k^{\to b_\mathrm{u}}(\Bar{h}_\mathrm{p}(\xi)) =\gamma^- k^{\to a}(\Bar{h}_\mathrm{p}(\xi))$, forms a boundary between regime (\MakeUppercase{\romannumeral 2}) and unbounded growth. This boundary marks the onset of positivity in the largest eigenvalue of the system described by the effective rates, which we  hereafter refer to uniquely as the effective eigenvalue. The endpoints of the critical line, $\xi_c$ and $\xi^*$, are located at $\Bar{h}=\sup \Bar{h}_\mathrm{R}$ and $\Bar{h} =0$, respectively, where $\Bar{h}_\mathrm{R}\in\{\Bar{h}\, |\, N_B(\Bar{h},\xi)=\max_{\Bar{h}'}N_B(\Bar{h}',\xi)\}$ denotes the destabilization boundary in regime (\MakeUppercase{\romannumeral 1}).
Assume now that the system state is initially in (\MakeUppercase{\romannumeral 2}) close to the critical line. If the load per bond increases, either through fluctuations or through external influence, $\Bar{h}$ will grow. 
However, the system cannot attain a steady state with an average stretch exceeding $\Bar{h}_\mathrm{p}$, as the onset of unbounded growth at this point reduces the average stretch below $\Bar{h}_\mathrm{p}$. This effect bounds the growth of $\Bar{h}$, thereby preventing rupture. Thus, while the divergent growth under load for $\xi_c\leq\xi<\xi^*$ mirrors the equilibrium instability, it enables perfect stability of the adhesion cluster.

\subsection{Broad range of perfect stabilization}
The condition for perfect stabilization can be visualized graphically by crossings of the numerically calculated functions $k^{\to b_\mathrm{u}}$ and $k^{\to a}$, which yield $\Bar{h}_{\mathrm{p}}$; see inset of Fig.~\ref{fig:phase-plot}\textsf{C}. A broad range $[\xi_c,\xi^*)$ is characterized by strong growth of $k^{\to b_\mathrm{u}}$ with respect to $k^{\to a}$. This range depends on the system parameters but is typically broad for physiologically relevant scales (Supplementary Information, Fig.~S2 and S3). Particularly important are the parameters controlling the accessibility (e.g., $\epsilon_\mathrm{f}$ and $\epsilon_\mathrm{b}$) and the mechanosensitivity ($\Delta$ and $\ell_\mathrm{b}$) of the bound molecules. In the limit of force-insensitive folding transitions $\Delta\to0$, we find $\xi_c\to\xi^*$, highlighting the crucial feature of force-sensitive coupling.

\subsection{Dynamic loading}
\begin{figure}[H]
    \centering
\includegraphics[width=1.\linewidth]{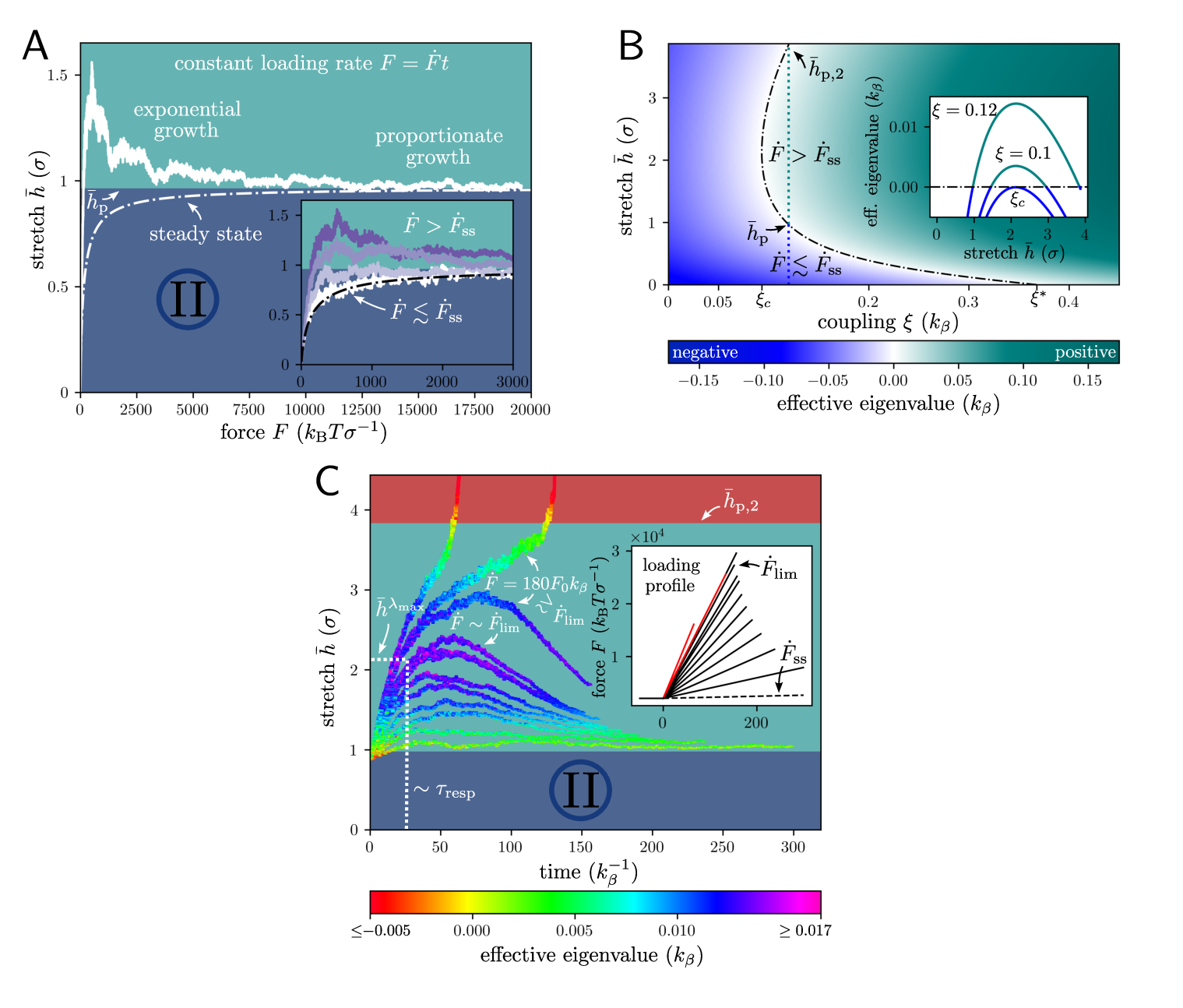} 
    \caption{Model behavior under dynamic loading. (\textsf{A})~A sample trajectory for $\xi=0.12k_{\beta}$, with constant loading rate $\Dot{F}=15F_0k_{\beta}$ ($F_0=k_\mathrm{B}T\sigma^{-1}$) starting from $F=0$. The system passes beyond $\Bar{h}_\mathrm{p}$, undergoes exponential growth followed by proportional growth with force. Inset: sample trajectories with $\Dot{F}\in\{1,5,11,15\}F_0k_{\beta}$. (\textsf{B})~Largest effective eigenvalue distribution estimated from the mean-field solutions. For $\Dot{F}>\Dot{F}_{\mathrm{ss}}$, the system can develop a positive effective eigenvalue. Inset shows cuts for three values of the control parameter $\xi$. (\textsf{C})~ Sample trajectories with $\Dot{F}\in \{20,40,\ldots,120,140,150,170,180,220\}F_0k_{\beta}$ showing transient exponential growth and instabilities for force ramps starting at a steady state near $\Bar{h}_\mathrm{p}$ with $\xi=0.12k_{\beta}$. Effective eigenvalue is computed from simulations, and its mean agrees with the result in (\textsf{B}) (see also Supplementary Information, Fig.~S7). Inset: loading profiles for each plotted trajectory. For $\Dot{F}>\Dot{F}_{\mathrm{lim}}\approx170F_0k_{\beta}$, the system can become unstable (red lines).}
    \label{fig:dynamic}
\end{figure}

To go beyond the static load assumption made so far, we next study linear force ramps. In the simulations, force ramps are approximated by incrementally increasing the force by an amount $\mathrm{d}F$ at fixed time intervals $\mathrm{d}t$, so that an effective loading rate is given by $\dot{F}=\mathrm{d}F/\mathrm{d}t$ (Supplementary Information). For intrinsic binding rates relevant in cell adhesions $k_\beta\sim 10^{-1}-10^1\,\mathrm{s}^{-1}$~\citep{bihr2014association}, loading rates $\sim 10^0-10^2\,k_\mathrm{B}T\sigma^{-1}k_\beta\sim 10^{-1}-10^3\,\mathrm{pN}\mathrm{s}^{-1}$ are consistent with physiologically relevant scales \citep{evans2007forces,jo2024determination}. For regimes (\MakeUppercase{\romannumeral 1}) and (\MakeUppercase{\romannumeral 2}), we find a limiting loading rate $\dot{F}_{\mathrm{ss}}$, below which the system's transient response closely approximates the steady-state limit independently of $\xi$, where, numerically, $\Dot{F}_{\mathrm{ss}}\approx 1.7k_\mathrm{B} T \sigma^{-1}k_{\beta}$ (Supplementary Information, Fig.~S5). For $\Dot{F}>\Dot{F}_{\mathrm{ss}}$, in regime (\MakeUppercase{\romannumeral 2}), the system can overshoot beyond $\Bar{h}_\mathrm{p}$, developing a positive effective eigenvalue before either stabilizing or rupturing. During a stable overshoot, the average stretch decreases due to cluster growth and eventually reaches a value close to the asymptote $\Bar{h}_\mathrm{p}$, where perfect stabilization is achieved, see sample trajectory in Fig.~\ref{fig:dynamic}\textsf{A}. The effective eigenvalue peaks at a positive value at average stretches of $\Bar{h}^{\lambda_{\mathrm{max}}}$ before returning to negative values at $\Bar{h}_{\mathrm{p},2}$, see Fig.~\ref{fig:dynamic}\textsf{B}.
Thus, the domain $[\Bar{h}_\mathrm{p},\Bar{h}_{\mathrm{p},2}]$ acts as a stabilizing stretch buffer against perturbations and fluctuations (Supplementary Information, Fig.~S4). Stable overshoots occur if loading rates are smaller than $\Dot{F}_{\mathrm{lim}} \approx\kappa N_B^\mathrm{i}(\Bar{h}^{\lambda_{\mathrm{max}}}-\Bar{h}^\mathrm{i})/\tau_{\mathrm{resp}}$, where $\tau_{\mathrm{resp}}$ characterizes a lag time resulting from the growth dynamics and $N_B^\mathrm{i}$ and $\Bar{h}^\mathrm{i}$ are the steady-state quantities before commencement of loading. Note that the value for $\Dot{F}_{\mathrm{lim}}$ is dependent on the initial state of the cluster. For loading proximate to the asymptote, as $\xi\to\xi_c$, $\Bar{h}_{\mathrm{p},2}-\Bar{h}_\mathrm{p}\to 0$ and so $\Dot{F}_{\mathrm{lim}}\to\Dot{F}_{\mathrm{ss}}$.

Figure~\ref{fig:dynamic}\textsf{C}~shows sample trajectories for which the force ramp starts from a steady state close to $\Bar{h}_{\mathrm{p}}$ with $\xi=0.12k_{\beta}>\xi_c$. For $\Dot{F}>\Dot{F}_{\mathrm{lim}}$, after the lag time $\sim\tau_{\mathrm{resp}}$, the positive effective eigenvalue decreases with increasing stretch, reducing the growth response despite rising forces. Consequently, the cluster average stretch can diverge. Note the broad range of loading rates that allow stabilization beyond the limit of quasi-static perfect stabilization $\Dot{F}_{\mathrm{lim}}\gg\Dot{F}_{\mathrm{ss}}$, see inset Fig.~\ref{fig:dynamic}\textsf{C} and Supplementary Information, Fig.~S6.

\subsection{Perfect stabilization in generic networks}

\begin{figure}[t!]
    \centering 
 \includegraphics[width=1.\linewidth]{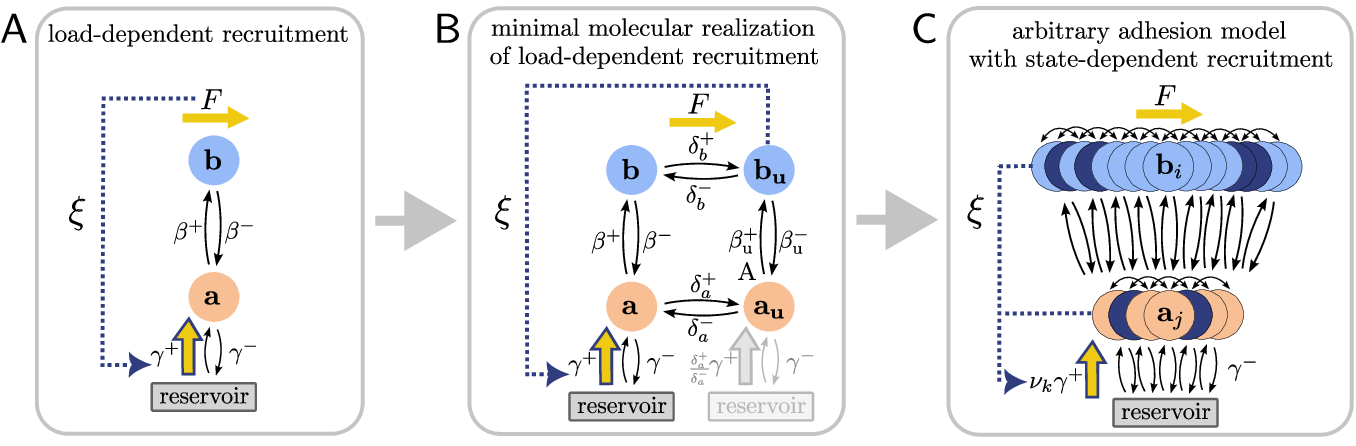}
    \caption{ Flow diagram of molecular adhesion models with load-dependent recruitment of increasing complexity. (\textsf{A}) Two-state model, where the recruitment rate is directly coupled to the applied force through coupling to the average stretch of the bound states. (\textsf{B}) The four-state model is a minimal implementation of load-dependent recruitment, where the force coupling is realized by coupling to the occupation number of the bound-unfolded state $b_\mathrm{u}$. Allowing the unbound-unfolded state $a_\mathrm{u}$ to exchange with the reservoir (shown in gray) results in the breakdown of limited self-stabilization established in Ref.~\cite{braeutigam2022generic}, but not the breakdown of perfect stabilization (see Supplementary Information). (\textsf{C}) A general architecture of an adhesion model with bound and unbound states with occupation numbers coupled to the reservoir can also result in the existence of a perfect stabilization regime.}
    \label{fig:generic}
\end{figure}

In the aforementioned four-state model, the unfolded bound state acts as a force sensor, which is critical for the emergence of a perfect stabilization regime. The same mechanism can be isolated in an even simpler two-state model of a bound and a reservoir-connected unbound state, in which the recruitment rate is coupled directly to the first moment of the bound-state stretch distribution, $\int_{\mathbb{R}^+}n_b(h)h\,\mathrm{d}h= N_b \Bar{h} = F/\kappa$ under force balance (see Fig.~\ref{fig:generic}\textsf{A} and Supplementary Information). The four-state model, shown again in Fig.~\ref{fig:generic}\textsf{B}, can therefore be viewed as a minimal molecular implementation of the same load-dependent recruitment principle.

To generalize beyond the four-state model, we consider a connected network with $x$ bound states $b_i$ and $y$ unbound states $a_j$, depicted in Fig.~\ref{fig:generic}\textsf{C}, with arbitrary stretch-dependent transition rates consistent with the equilibrium cycle conditions, so that the cluster remains passive under zero applied force. We assume mechanical relaxation between state transitions occurs instantaneously, so that the applied force is balanced between the bound states. The states $\{a_i|i\in \mathcal{I}\}$ are connected to a common reservoir, where $|\mathcal{I}|\leq y$, which are coupled to internal states $\{p_i|i\in \mathcal{J}\}$, where $|\mathcal{J}|\leq x+y$, i.e., coupled states may be both bound and unbound states. To satisfy the Kolmogorov cycle conditions, each coupled reservoir recruitment rate in the network must couple to the same set of states with appropriate scaling $\nu_i$, that is, $\gamma^+_i(\xi)= \nu_i(\gamma_0^+ + \xi \sum_{j\in\mathcal{J}}N_{p_j})$ for $i\in\mathcal{I}$. For instance, when connecting the four-state model to the reservoir through the additional state $a_\mathrm{u}$, the corresponding recruitment rate must be scaled as $\frac{\delta_a^+}{\delta_a^-}\gamma^+(\xi)=e^{-\epsilon_\mathrm{f}/k_\mathrm{B}T}\gamma^+(\xi)$ (see Fig.~\ref{fig:generic}\textsf{B}).

Owing to the linearity of the transport and state-occupation equations, the normalized state-occupation distributions are independent of $\xi$ (Supplementary Information). Consequently, effective rates as defined in Eq.~(\ref{effective_rates}) will be independent of $\xi$. Thus, the stationary nonequilibrium occupation number of a state $q$ obeys
\begin{equation}
    {N}_q(\Bar{h},\xi) = \frac{\gamma_0^+ k^{\to q}(\Bar{h})}{\gamma^-K^{\to a}(\Bar{h})-\xi K^{\to p}(\Bar{h})}, 
\end{equation}
where $K^{\to a}(\bar{h}):=(\sum_{i\in \mathcal{I}}\nu_i)^{-1} \sum_{i\in \mathcal{I}} k^{\to a_i}(\Bar{h})$ and $K^{\to p}(\Bar{h}):=\sum_{j\in\mathcal{J}}k^{\to p_j}(\Bar{h})$. These functions capture the effective transitions into the reservoir-connected states and the coupled states, respectively. Defining the function $\xi_\mathrm{div}(\bar{h}):=\gamma^-\frac{K^{\to a}}{K^{\to p}}(\bar{h})$, the necessary and sufficient condition for the existence of a perfect stabilization regime is
\begin{equation}\label{eq:general_cond}
   \min_{\bar{h}\in\mathcal{H}_\mathrm{acc},\bar{h}>0}\xi_\mathrm{div}(\bar{h})< \xi_\mathrm{div}(0)\equiv \xi^*,
\end{equation}
where $\mathcal{H}_\mathrm{acc}$ denotes the mechanically accessible range of the system. Without such a constraint, the above condition could be satisfied for a system where the coupled states only become mechanosensitive for large $\bar{h}$ after full rupture. For instance, in the four-state model, for large $\ell_\mathrm{b}$ and $\Delta$, the system fully destabilizes before the average stretch is sufficiently large for the unfolding events to dominate (see Supplementary Information).

Around equilibrium, $\xi_\mathrm{div}(\bar{h}) = \xi^*(1- \mathcal{M}\bar{h}) + \mathcal{O}(\bar{h}^2)$, where $\mathcal{M}\equiv\frac{\mathrm{d}}{\mathrm{d}\bar{h}}\ln \frac{K^{\to p}}{K^{\to a}} \big |_0$, which captures the mechanical sensitivity of transitions into the coupled states relative to the reservoir-connected states. Therefore, when
\begin{equation}\label{eq: local_cond}
    \frac{\mathrm{d}}{\mathrm{d}\bar{h}}\ln \frac{K^{\to p}}{K^{\to a}} \bigg |_0 >0,
\end{equation}
for an arbitrarily small interval below $\xi^*$, the system can achieve perfect stabilization. To further interpret the above condition, recall that $k^{\to q}\equiv \sum_{\mathcal{T}\in\mathcal{T}^q} \prod_{(p',q')\in \mathcal{T}}\Bar{k}^{p'\to q'}$, which denotes the sum of products of rates $\Bar{k}^{p'\to q'}$ associated with the edge $(p',q')$ that belongs to a spanning tree $\mathcal{T}$ belonging to the set of spanning trees $\mathcal{T}^q$ of the network graph with edges directed towards $q$. For convenience, we further write $ W_{\mathcal{T}} \equiv \prod_{(p',q')\in \mathcal{T}}\Bar{k}^{p'\to q'}$, so that $k^{\to q}= \sum_{\mathcal{T}\in\mathcal{T}^q} W_{\mathcal{T}}$. It follows that 
\begin{equation}
      \frac{\mathrm{d}}{\mathrm{d}\bar{h}}W_{\mathcal{T}}\bigg|_0 = W_{\mathcal{T}}^0S_{\mathcal{T}},
\end{equation}
where $S_{\mathcal{T}}\equiv\sum_{(p',q')\in \mathcal{T}}s^{p'\to q'}$, with $s^{p'\to q'}=  \frac{\mathrm{d}}{\mathrm{d}\bar{h}}\ln \Bar{k}^{p'\to q'} \big|_0$, and $W_{\mathcal{T}}^0\equiv W_{\mathcal{T}}(0)$. Now, let $\mathscr{T}_a = \cup_{i\in\mathcal{I}}\mathcal{T}^{a_i}$ and $\mathscr{T}_p = \cup_{j\in\mathcal{J}}\mathcal{T}^{p_j}$ denote the set of trees rooted in the reservoir-connected and coupled states, respectively, so that $K^{\to a}= (\sum_{i\in \mathcal{I}}\nu_i)^{-1}  \sum_{\mathcal{T}\in \mathscr{T}_a}W_{\mathcal{T}}$ and $K^{\to p}=  \sum_{\mathcal{T}\in \mathscr{T}_p}W_{\mathcal{T}}$. Then 
\begin{equation}
    \begin{split}
         \frac{\mathrm{d}}{\mathrm{d}\bar{h}} \ln K^{\to a} \bigg |_0 &= \frac{\sum_{\mathcal{T}\in \mathscr{T}_a}W_{\mathcal{T}}^0S_{\mathcal{T}}}{\sum_{\mathcal{T}\in \mathscr{T}_a}W_{\mathcal{T}}^0} \\
         &=: \langle  S_a \rangle_{0,W},
    \end{split}
\end{equation}
and the analogous relation holds for $\langle  S_p \rangle_{0,W}$. This expression is an equilibrium-weighted average of the sums of the logarithmic stretch sensitivities of the effective rates across the edges of spanning trees directed toward the reservoir-connected and coupled states, respectively. Therefore, Eq.~(\ref{eq: local_cond}) is equivalent to
\begin{equation}\label{eq:S-cond}
    \langle  S_p \rangle_{0,W} > \langle  S_a \rangle_{0,W}.
\end{equation}
That is, a perfect-stabilization interval opens locally below $\xi^*$ when the equilibrium-weighted logarithmic mechanosensitivity of directed reaction trees rooted in the coupled states exceeds that of trees rooted in the reservoir-connected states. Importantly, this is a network property rather than solely a property of the coupled state itself: a coupled state may inherit mechanosensitivity through force-sensitive upstream routes. For example, in the four-state model, recruitment coupled to $a_\mathrm{u}$ can produce perfect stabilization, since trees rooted in $a_\mathrm{u}$ include contributions containing the force-sensitive route through $b_\mathrm{u}$, such as $b\to b_\mathrm{u}\to a_\mathrm{u}$ (see Supplementary Information). Moreover, by approximating the deformation of the equilibrium Boltzmann stretch distributions to first order in $\Bar{h}$ as a shift in the mean by $\Bar{h}$, the sensitivities $s^{p'\to q'}$ can typically be formulated in terms of the model parameters. For example, for the four-state model, Eq.~(\ref{eq:S-cond}) breaks down as the unfolding length $\Delta\to 0$ (see Supplementary Information).

Note, however, that this local condition is not necessary, since the system's mechanosensitivity may ``turn on'' further out of equilibrium (see Supplementary Information), while the condition in Eq.~(\ref{eq:general_cond}) is necessary and sufficient. Therefore, in direct analogy with the four-state model studied above, appropriate coupling to force-sensitive states can drive load-dependent recruitment of molecules and lead to perfect stabilization in a broad class of networks representing complex biomolecular adhesion clusters.

\section{Discussion}
Inspired by the well-established dependence of focal-adhesion size on mechanical load, we propose a general model of biomolecular adhesion in which the recruitment of molecules to the cluster is governed by the internal states of molecules already present within the cluster. Using a versatile modeling framework for state-dependent recruitment, we show how biomolecular adhesion clusters can achieve perfect stability by adjusting their size to arbitrarily large stationary loads within the idealized model. Perfect stabilization is based on a growth instability that emerges in nonequilibrium systems with state-dependent recruitment of molecules. A parameter variation shows that the perfect-stabilization interval can remain broad over physiologically relevant parameter ranges (Supplementary Information, Figs.~S3 and S9). Although we use a minimal model, we aimed to employ physiologically relevant parameter values (Supplementary Information, Table~S2). Using intrinsic binding rates of $k_\beta \simeq 10^{-1}-10^1\,\mathrm{s}^{-1}$, as reported for cell adhesions~\citep{bihr2014association}, the resulting timescales for perfect stabilization, $\xi_c^{-1} \simeq 10 k_\beta^{-1} \simeq 10^{0}-10^{2}\,\mathrm{s}$, are plausible. For instance, vinculin binding to unfolded talin occurs within approximately one second~\citep{tapia2020direct}. Load-dependent recruitment of adhesion molecules, which could potentially enable perfect stabilization, is mechanistically distinct from catch-bond behavior~\citep{de2018general,novikova2021evolving}. It is also distinct from previously described mechanisms of limited self-stabilization~\citep{braeutigam2022generic,braeutigam2024clutch}, because it enables a regime of perfect stabilization in which no finite stationary load exists above which the adhesion cluster becomes unstable; see regime~(\MakeUppercase{\romannumeral 2}) in Fig.~\ref{fig:sim-full}.

Predictions arising from state-dependent recruitment of molecules can be used to test for this mechanism in different types of cellular adhesion sites. State-dependent recruitment can produce an increase in the number of adhesion molecules with the total force acting on the adhesion cluster. Although considerable work has established a correlation between the size of maturing focal adhesions and force~\citep{balaban2001force,tan2003cells,goffin2006focal,stricker2011spatiotemporal,trichet2012evidence,bergert2016confocal}, it remains unclear how focal adhesions adapt under very large forces. Experiments could directly examine the limiting strength of focal adhesions under shear stress; see, for example, Ref.~\citep{paddillaya2019biophysics} and the references therein. Such experiments would have to carefully control for confounding factors that could influence the mechanical response of the focal adhesion, including stresses transmitted through the broader cytoskeleton~\citep{verma2015flow}. In the regime of perfect stabilization, the model predicts linear growth of the number of adhesion molecules with force, without the high-load saturation that would ultimately lead to rupture. At the same time, the average load per bond approaches a plateau. Furthermore, the model predicts that reducing the rate of force-sensitive recruitment below a critical threshold, $\xi < \xi_c$, restores a finite rupture force, as illustrated in Fig.~\ref{fig:sim-full}\textsf{B}. Conversely, when the recruitment coupling exceeds the threshold $\xi^{*}$, growth is no longer controlled by the force dependence of the recruitment mechanism. Instead, focal-adhesion growth would be limited by factors not included in our model, such as diffusion, molecular availability, spatial constraints, and biochemical regulation. In this regime, the model does not predict a linear relationship between focal-adhesion size and force; rather, it predicts large, force-insensitive adhesion clusters over a range of applied loads.

Genetic and pharmacological perturbation experiments demonstrate that force-dependent reinforcement and maturation of focal adhesions (FAs) involve molecular-state transitions and signaling pathways mediated by adaptor proteins such as talin, vinculin, and paxillin, as well as by FAK and mDia1. Although the transmission of mechanical forces can regulate adhesion-complex composition and phosphotyrosine signaling even in the absence of mechanically regulated talin rod subdomains~\citep{rahikainen2019talin}, force-induced talin unfolding and the resulting exposure of cryptic binding sites within the rod subdomains have been shown to be important for focal-adhesion reinforcement~\citep{del2009stretching,goult2013riam,austen2015extracellular,elosegui2016mechanical}. Stepwise destabilization of the talin R3 subdomain results in the formation of large, talin-rich adhesion clusters with increased molecular turnover~\citep{rahikainen2017mechanical}. This observation qualitatively resembles the model prediction that lowering the force threshold for transitions into the unfolded state increases the occupation of recruitment-promoting states and thereby enhances adhesion-cluster growth.

Vinculin mechanically reinforces the linkage between talin and the actin cytoskeleton, while vinculin binding stabilizes the unfolded state of talin~\citep{del2009stretching,ciobanasu2014actomyosin}. An allosteric activation mechanism further regulates talin-vinculin interactions during the transition from initial binding to adhesion maturation~\citep{franz2023allosteric}. Expression of constitutively active vinculin generates large, reportedly force-independent FAs~\citep{carisey2013vinculin,atherton2015vinculin}. Tentatively, the formation of a vinculin-mediated cross-link between talin and actin could be represented in our model by a change in the effective affinity of the bound states. Because $\xi^{*}$ decreases with increasing bound-state affinity in the model, constitutively active vinculin could, in principle, shift the system into the regime $\xi>\xi^{*}$ by lowering $\xi^{*}$ rather than by directly increasing $\xi$. In this regime, adhesion-cluster growth is predicted to occur even in the absence of force.

Experiments in which the loading rate applied to FAs was controlled have revealed rate-dependent adhesion growth and reinforcement, with cytoskeletal softening limiting reinforcement at high loading rates~\citep{andreu2021force}. The present model predicts adhesion growth proportional to force during sufficiently slow loading, but a transient overshoot beyond the quasi-static stabilization point and an eventual loss of stabilization at sufficiently high loading rates. The critical loading rate increases with the initial size of the adhesion cluster.

Measurements of actin-flow velocities could, in principle, be compared with predictions for the drift velocity $v$ in focal-adhesion models~\citep{gardel2008traction,sabass2010modeling,oakes2012tension}. However, such a comparison would require an extended model that explicitly represents actin polymerization, cytoskeletal remodeling, and focal-adhesion dynamics. For the parameter values used here (Supplementary Information, Table~S2), the present reduced model predicts steady-state drift velocities of $v \sim \sigma k_\beta \simeq 10^{-1}-10^{1}\,\mathrm{nm\,s^{-1}}$, as shown in Fig.~\ref{fig:sim-full}\textsf{A}.

From a materials-science perspective, our model offers a conceptual blueprint for nonrupturing, bioinspired junctions with potential engineering applications. Materials incorporating the self-stabilizing mechanism described here could combine softness with high load-bearing capacity, renew themselves through molecular turnover, and permit chemical or mechanical control of their strength. Thus, the concept of perfect adhesion stability provides another example of how biological principles can inspire soft materials with unusual mechanical properties.

Overall, this work demonstrates that perfect adhesion stability is theoretically possible and can arise in models of biomolecular adhesion in which increasing mechanical load enhances the recruitment of adhesion molecules. The framework presented here could be extended to represent the molecular biology of focal adhesions in greater detail. Future work could explicitly describe multimolecular interactions among selected components while retaining a framework that respects thermodynamic constraints. State-dependent recruitment of molecules can naturally generate nonequilibrium growth instabilities~\citep{marehalli2024thermodynamics,PhysRevE.109.064153}, which, under the conditions identified here, can produce perfect stabilization. Important constraints that may limit perfect stabilization include finite molecular pools, spatial limitations on adhesion size, diffusion-limited recruitment, and assembly or disassembly processes that compete with recruitment. Incorporating these effects is a topic for future work.

\section{Methods}
\subsection{Transition rates}
The stretch-dependent rates are constructed such that Kolmogorov's cycle condition is respected~\citep{kolmogoroff1936theorie}. This condition means here that when a single molecule goes through a cycle of states in absence of driving forces, e.g., $a\to b\to b_\mathrm{u}\to a_\mathrm{u} \to a$, the product of transition rates in one direction equals the product of reverse transition rates as
\begin{equation}
1 = \frac{\beta^+(h) \delta_b^+(h) \beta_\mathrm{u}^-(h-\Delta)\delta_a^-}{\beta^-(h) \delta_b^-(h-\Delta) \beta_\mathrm{u}^+(h-\Delta)\delta_a^+}.
\label{eq:rates:kolmogorov}
\end{equation}
This thermodynamic constraint follows from the principle of microscopic reversibility and ensures that our molecules do not act as molecular motors, but rather as passive adhesions. See Supplementary Information for further details.
\subsection{Simulations}
The stochastic simulations were performed using the Gillespie algorithm~\citep{gillespie1976general,gillespie1977exact}. After each event, the individual bound molecule stretches are updated to ensure force balance. For binding events, the molecules are initialized with stretches drawn from the unbound stretch distribution $\eta(h)\sim\mathcal{N}(0,\sigma^2)$. See Supplementary Information for further details.

\section{Author Contributions}
A.F.B. and B.S. designed and performed the research. J.M. developed an initial model and conducted numerical analysis. A.F.B. and B.S. discussed results and wrote the manuscript.

\section{Acknowledgments}
We acknowledge funding by the European Research Council (ERC) under the European Union's Horizon 2020 research and innovation program (BacForce, G.A. No. 852585) and by the Deutsche Forschungsgemeinschaft (DFG, German Research Foundation), Project No. 492014049. The authors declare that they have no competing interests.

\section{Data availability}
The code used to perform the numerical simulations
and generate the results in this work is publicly available
at \url{https://github.com/AFBurnet/State_dependent_recruitment_biomolecular_adhesion_model}.

\bibliographystyle{unsrtnat}
\bibliography{bibliography}

\end{document}